\magnification=\magstep1
\baselineskip=18pt
\input epsf

\centerline {\bf Wilson Fermions at finite temperature}
\bigskip
\centerline {Michael Creutz}
\centerline {Physics Department}
\centerline {Brookhaven National Laboratory, Upton, NY 11973}
\medskip
\centerline {creutz@bnl.gov}
\vskip 1.5in
\centerline {Abstract}

I conjecture on the phase structure expected for lattice gauge theory
with two flavors of Wilson fermions, concentrating on large values of
the hopping parameter.  Numerous phases are expected, including the
conventional confinement and deconfinement phases, as well as an Aoki
phase with spontaneous breaking of flavor and parity and a large
hopping phase corresponding to negative quark masses.

\vfill\eject

In this talk I conjecture on the rather rich phase structure expected
for lattice gauge theory with Wilson fermions, paying particular
attention to what happens for large hopping parameter.  I consider
both zero and non-zero temperature.  I restrict myself to the
standard hadronic gauge theory of quarks interacting with non-Abelian
gluons.  I leave aside issues related to electromagnetism and weak
interactions, both of which also raise fascinating issues for lattice
field theory.

The parameters of the strong interactions are the quark masses.  I
implicitly include here the strong CP violating parameter $\theta$, as
this can generally be rotated into the mass matrix [1].  The quark
masses are in fact the only parameters of hadronic physics, the strong
coupling being absorbed into the units of measurement via the
phenomenon of dimensional transmutation [2].

For the purposes of this talk, I take degenerate quarks at $\theta=0$;
so, I can consider only a single mass parameter $m$.  I discuss only
the two flavor case, as this will make some of the chiral symmetry
issues simpler.  I also will treat the theory at finite temperature,
$T$, introducing another variable.  Finally, as this is a lattice
talk, I introduce the lattice spacing $a$ as a third parameter.

On the lattice with Wilson fermions the three parameters $(m,T,a)$ are
usually replaced with $\beta$, representing the inverse bare lattice
coupling squared, the fermion hopping parameter $K$, and the number of
time slices $N_t$.  The mapping between $(m,T,a)$ and $(\beta,K,N_t)$
is non-linear, well known, and not the subject of this talk.

Note that in considering the structure of the theory in either of
these sets of variables, I am inherently talking about finite lattice
spacing $a$.  Thus this entire talk is about lattice artifacts.

I start with the $(\beta,K)$ plane at zero temperature, and defer how
this is modified at finite temperature.  The $\beta$ axis with $K=0$
represents the pure gauge theory of glueballs.  This is expected to be
confining without any singularities at finite $\beta$.  The line of
varying $K$ with $\beta=\infty$ represents free Wilson fermions [3].
Here, with conventional normalizations, the point $K={1\over 8}$ is
where the mass gap vanishes and a massless fermion should appear in
the continuum limit.  The full interacting continuum limit should be
obtained by approaching this point from the interior of the
$(\beta,K)$ plane.

While receiving the most attention, this point $K={1\over 8}$ is not
the only place where free Wilson fermions lose their mass gap.  At
$K={1\over 4}$ four doubler species become massless. Also formally at
$K=\infty$ six more doublers loose their mass.  (Actually, a more
natural variable is ${1\over K}$.)  The remaining doublers occur at
negative $K$.

The $K$ axis at vanishing $\beta$ also has a critical point where the
confining spectrum appears to develop massless states.  Strong
coupling arguments as well as numerical experiments place this point
somewhere near $K={1\over 4}$, but this is probably not exact.  The
conventional picture connects this point to $(\beta=\infty, K={1\over
8})$ by a phase transition line representing the lattice version of
the chiral limit.

\topinsert
{
\epsfxsize .7\hsize
\centerline {\epsfbox{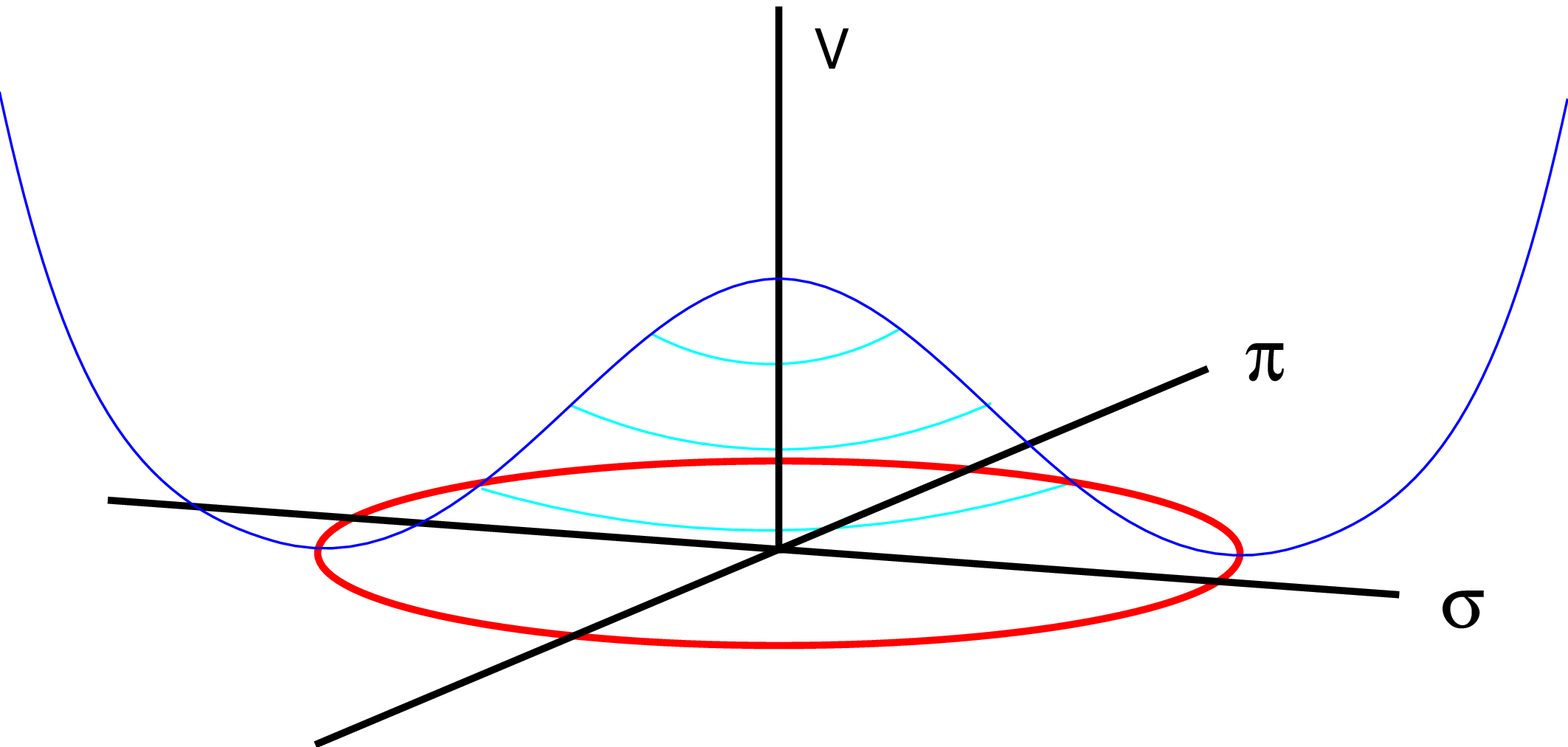}}
Fig.~(1) The effective potential in the canonical sigma model.
\bigskip
}
\endinsert

Now I move ever so slightly inside the $(\beta,K)$ plane from the
point $(\infty, {1\over 8})$.  This should take us from free quarks to
a confining theory, with mesons, baryons, and glueballs being the
physical states.  Furthermore, when the quark is massless, we should
have chiral symmetry.  Considering here the two flavor case, this
symmetry is nicely exemplified in a so called ``sigma'' model, with
three pion fields and one sigma field rotating amongst themselves.
Defining the fields
$$
\eqalign{
&\sigma=\overline\psi\psi\cr
&\vec\pi=i\overline\psi\gamma_5\vec\tau\psi\cr
} \eqno (1)
$$
I consider constructing an effective potential.  For massless quarks
this is expected to have the canonical sombrero shape stereotyped by
$$
V\sim\lambda(\sigma^2+\vec\pi^2-v^2)^2 
\eqno (2)
$$
and illustrated schematically in Fig.~(1).  The normal {\ae}ther is
taken with an expectation value for the sigma field
$\langle\sigma\rangle\sim v$.  The physical pions are massless
goldstone bosons associated with slow fluctuations of the {\ae}ther
along the degenerate minima of this potential.
 
As I move up and down in $K$ from the massless case near ${1\over 8}$,
this effective potential will tilt in the standard way, with the sign
of $\langle\sigma\rangle$ being appropriately determined.  The role of
the quark mass is played by the distance from the critical hopping,
$m_q\sim K_c-K$ with $K_c\sim {1\over 8}$.  At the chiral point there
occurs a phase transition, of first order because the sign of
$\langle\sigma\rangle$ jumps discontinuously.  At the transition point
there are massless goldstone pions representing the spontaneous
symmetry breaking.  With an even number of flavors the basic physics
on each side of the transition is the same, since the sign of the mass
term is a convention reversable via a chiral rotation.  For an odd
number of flavors the sign of the mass is significant because the
required rotation involves the $U(1)$ anomaly and is not a good
symmetry.  This is discussed in some detail in my recent paper,
Ref. [1].  For the present discussion I stick with two flavors.

A similar picture should also occur near $K={1\over 4}$, representing
the point where a subset of the fermion doublers become massless.
Thus another phase transition should enter the diagram at $K={1\over
8}$.  Similar lines will enter at negative $K$ and further complexity
occurs at $K=\infty$.  For simplicity, let me concentrate only on the
lines from $K={1\over 8}$ and ${1\over 4}$.

Now I delve a bit deeper into the $(\beta,K)$ plane.  The next
observation is that the Wilson term separating the doublers is
explicitly not chiral invariant.  This should damage the beautiful
symmetry of our sombrero.  The first effect expected is a general
tilting of the potential.  This represents an additive renormalization
of the fermion mass, and appears as a beta dependent motion of the
critical hopping away from ${1\over 8}$.  Define $K_c(\beta)$ as the first
singular place in the phase diagram for increasing $K$ at given
$\beta$.  This gives a curve which presumably starts near $K={1\over
4}$ at $\beta=0$ and ends up at ${1\over 8}$ for infinite $\beta$.

Up to this point I have only reviewed standard lore.  Now I continue
to delve yet further away from the continuum chiral point at
$(\beta,K)=(\infty,{1\over 8})$.  Then I expect the chiral symmetry
breaking of the Wilson term to increase and become more than a simple
tilting of the Mexican hat.  I'm not sure to what extent a multipole
analysis of this breaking makes sense, but let me presume that the
next effect is a quadratic warping of our sombrero, i.e.  a term
something like $\alpha \sigma^2$ appearing in the effective sigma
model potential.  This warping cannot be removed by a simple mass
renormalization.

There are two possibilities.  This warping could be upward or downward
in the $\sigma$ direction.  Indeed, which possibility occurs can
depend on the value of $\beta$.

Consider first the case where the warping is downward, stabilizing the
sigma direction for the {\ae}ther.  At the first order chiral
transition, this distortion gives the pions a small mass.  The
transition then occurs without a diverging correlation length.  As
before, the condensate $\langle \sigma \rangle$ jumps discontinuously,
changing its sign.  However, the conventional approach of
extrapolating the pion mass to zero from measurements at smaller
hopping parameter will no longer yield the correct critical line.  The
effect of this warping on the potential is illustrated in Fig.~(2).

\topinsert
{
\epsfxsize .7\hsize
\centerline {\epsfbox{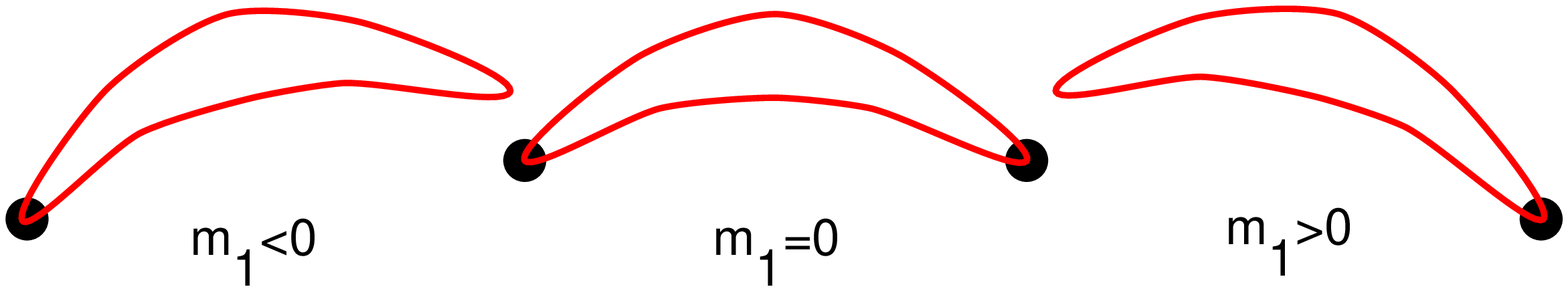}}
Fig.~(2) The effect of a downward warping of the effective potential.
The curve represents the warped bottom of the sombrero potential.
Here $m_1$ represents the distance from the critical point.
The solid circles represent possible states of the {\ae}ther.
The phase transition now occurs without a diverging correlation 
length.
\bigskip
}
\endinsert

A second possibility is for the warping to be in the opposite
direction, destabilizing the $\sigma$ direction.  In this case we
expect two distinct phase transitions to occur as $K$ passes through
the critical region.  For small hopping we have our tilted potential
with $\sigma$ having a positive expectation.  As $K$ increases, this
tilting will eventually be insufficient to overcome the destabilizing
influence of the warping.  At a critical point, most likely second
order, it will become energetically favorable for the pion field to
acquire an expectation value, such a case being stabilized by the
upward warping in the sigma direction.  As $K$ continues to increase,
a second transition should appear where the tilting of the potential
is again sufficiently strong to give only sigma an expectation, but
now in the negative direction.  The effect of this upward warping on
the effective potential is illustrated in Fig.~(3).

Thus we expect our critical line to split into two, with a rather
interesting phase between them.  This phase has a non-vanishing
expectation value for the pion field.  As the latter carries flavor
and odd parity, both are spontaneously broken.  Furthermore, since
flavor is still an exact continuous global symmetry, when it is broken
Goldstone bosons will appear.  In this two flavor case, there are
precisely two such massless excitations.  If the transitions are
indeed second order, a third massless particle appears just at the
transition lines, and these three particles are the remnants of the
three pions from the continuum theory.  This picture of a parity and
flavor breaking phase was proposed some time ago by Aoki [4], who
presented evidence for its existance in the strong coupling regime.
This phase should be ``pinched'' between the two transitions, and
become of less importance as $\beta$ increases.  Whether the phase
might be squeezed out at a finite $\beta$ to the above first order
case, or whether it only disappears in the infinite $\beta$ limit is a
dynamical question as yet unresolved.

\topinsert
{
\epsfxsize .7\hsize
\centerline {\epsfbox{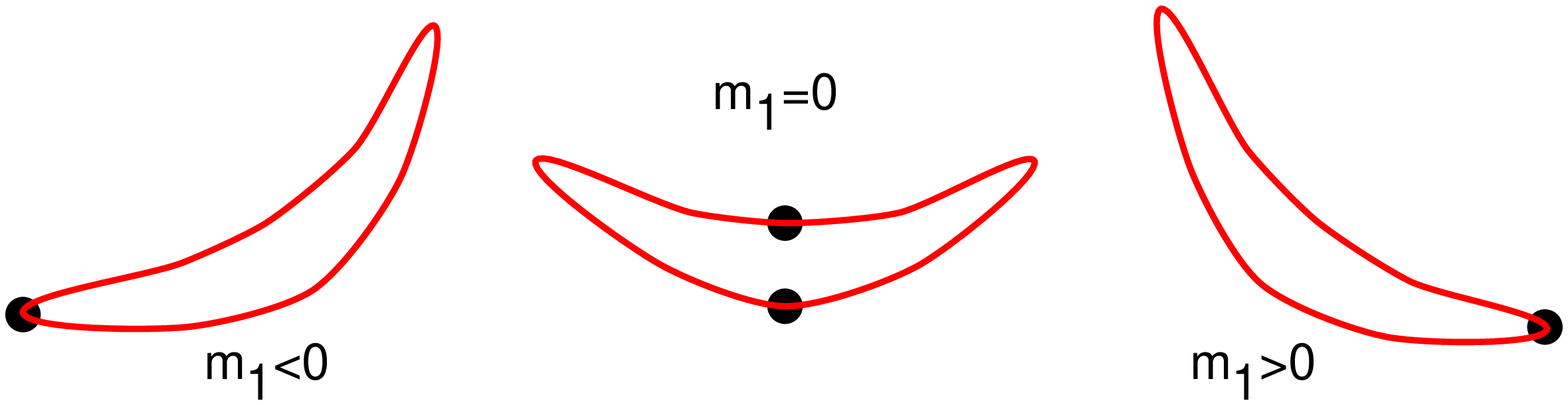}}
Fig.~(3) The effect of an upward warping of the effective potential.
Here $m_1$ represents the distance from $K_c(\beta)$.  Now there are
two phase transitions, with the intermediate phase having an
expectation for the pion field.
\bigskip
}
\endinsert

A similar critical line splitting to give a broken flavor
phase should also enter our phase diagram from
$(\beta,K)=(\infty,{1\over 4})$, representing the first set of
doublers.  Evidence from toy models [5] is that after this line
splits, the lower half joins up with the upper curve from the
$(\beta,K)=(\infty,{1\over 8})$ point.  In these models, there appears
to be only one broken parity phase at strong coupling.

Now let me go to finite temperature, or more precisely, finite $N_t$,
the number of sites in the temporal direction.  Along the $\beta$
axis, representing the pure glue theory, a deconfinement transition is
expected [6].  For an $SU(3)$ gauge group, this transition is expected
to be first order.  Turning on the fermion hopping, this transition
should begin to move in $\beta$, the first effect being an effective
renormalization of $\beta$ down toward stronger couplings.  In the
process, the transition may soften, and perhaps eventually turn into a
rapid crossover rather than a true singularity.  In any case, the
numerical evidence is for a single transition where both the Polyakov
line and the chiral symmetry order parameter undergo a rapid change.
The transition region should continue into the $(\beta,K)$ plane to
eventually meet the bulk transition line near $K_c(\beta)$ coming in
from strong coupling.

On the weak coupling side of the deconfinement transition, physics is
dramatically different.  Here as the quark mass goes to zero, we
expect chiral symmetry restoration in the thermal {\ae}ther.  In terms
of the effective potential, we expect only a single simple minimum.
Most importantly, we do not expect any singularity around zero quark
mass, with physics depending smoothly around the
$(\beta,K)=(\infty,{1\over 8})$ point.  In other words, we expect the
chiral transtion at small quark masses to be absorbed into the finite
temperature transition.  As the hopping continues to increase, the
$m\leftrightarrow -m $ symmetry of the continuum theory will play a
role, bouncing the deconfinement transition back towards larger
$\beta$ after $K$ passes $K_c$.

\topinsert
{
\epsfxsize .7\hsize
\centerline {\epsfbox{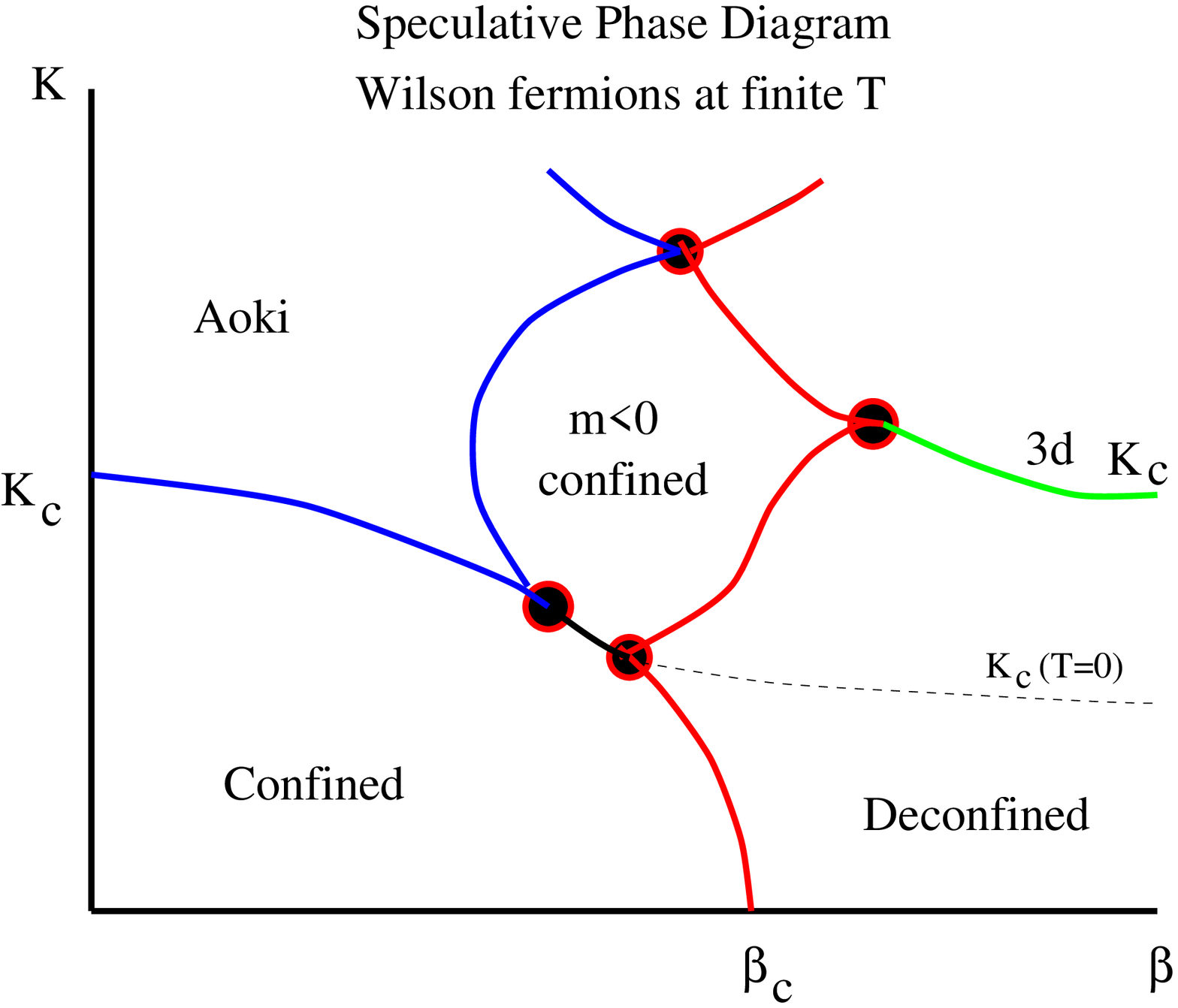}}
Fig. (4) The conjectured $(\beta,K)$ phase diagram for finite $N_t$.
\bigskip
}
\endinsert

What is less clear is what happens to the finite temperature line as
we continue further toward the chiral transitions of the doublers.
Here I conjecture that another transition line enters the picture.
For small $N_t$ the theory is effectively a three dimensional one,
which should have its own chiral transition, possibly somewhere
between $K={1\over 8}$ and $K={1\over 4}$.  Speculating that the
deconfinement transition bounces as well off of this line, but on the
opposite side, I arrive at the qualitative finite temperature phase
diagram sketched in Fig. (4).

To summarize the picture, at small $\beta$ and small $K$ we have the
usual low temperature confined phase.  Increasing $K$, we enter the
Aoki phase with spontaneous breaking of flavor and parity.  As $\beta$
increases, the Aoki phase pinches down into either a narrow point or a
single first order line, leading towards the free fermion point at
$(\beta,K)=(\infty,{1\over 8})$.  Before reaching that point, this
line collides with and is absorbed in the deconfinement transition
line.  The latter then bounces back towards larger $\beta$.  Above the
chiral line is a phase nearly equivalent physically with the usual
confined phase, just differing in the sign of the light quark masses.
Indeed, the only physical difference is via the lattice artifacts of
the doublers.  Finally, and most speculatively, there may be a three
dimensional chiral line coming in from large $\beta$ which reflects
the deconfinement transition back to meet the doubler chiral line
heading towards $(\beta,K)=(\infty,{1\over 4})$.

This diagram is wonderfully complex, probably incomplete, and may take
some time to map out.  Given the results presented by Ukawa at this
meeting [7], it appears that we may as yet be at too small a value of
$N_t$ for the negative mass confined phase to have appeared.  As a
final reminder, this entire discussion is of lattice artifacts, and
other lattice actions, perhaps including various ``improvements,''
will look dramatically different.

\bigskip
\noindent{\bf ACKNOWLEDGMENT}

This manuscript has been authored under contract number
DE-AC02-76CH00016 with the U.S.~Department of Energy.  Accordingly,
the U.S.~Government retains a non-exclusive, royalty-free license to
publish or reproduce the published form of this contribution, or allow
others to do so, for U.S.~Government purposes.  

\bigskip
\centerline{REFERENCES}
\parindent=0pt

1.  For a recent discussion of this old topic, see M. Creutz,
Phys. Rev. D52, 2951 (1995).

2. S. Coleman and E. Weinberg, Phys. Rev. D7 1888 (1973).

3. K.~Wilson, in {\sl New Phenomena in Subnuclear Physics}, 
Edited by A. Zichichi (Plenum Press, NY, 1977), p. 24. 

4. S. Aoki, Nucl. Phys. B314, 79 (1989); S.~Aoki and A.~Gocksch, 
Phys.~Rev. D45, 3845 (1992).

5. S.~Aoki, S.~Boetcher, and A.~Gocksch, Phys.~Lett. B331, 157
  (1994); K.~Bitar and P.~Vranas, Phys.~Rev. D50, 3406 (1994); Nucl.~Phys.
  B, Proc.~Suppl. 34, 661 (1994).

6. For a review see R. Gavai, in {\sl Quantum Fields on the
Computer}, M. Creutz, ed., p.~51 (World Scientific, 1992).

7. S.~Aoki, A.~Ukawa, and T.~Umemura, Phys.~Rev.~Lett.~76, 873 (1996).

\vfill\eject\bye